\documentclass[prb,twocolumn,showpacs,superscriptaddress]{revtex4}
\usepackage{graphicx}
\usepackage{dcolumn}
\usepackage{bm}

\begin{document}
\title{Stability of metallic  
stripes in the extended one-band Hubbard model}
\author{G. Seibold}
\affiliation{Institut f\"ur Physik, BTU Cottbus, PBox 101344, 
         03013 Cottbus, Germany}
\author{J. Lorenzana}
\affiliation{Center for Statistical
Mechanics and Complexity, INFM, Dipartimento di Fisica,
Universit\`a di Roma La Sapienza, P. Aldo Moro 2, 00185 Roma, Italy}
\date{\today}

\begin{abstract}
Based on an unrestricted Gutzwiller approximation (GA) we investigate 
the stripe orientation and periodicity in an extended  one-band
Hubbard model. 
A negative ratio between next-nearest and nearest neighbor hopping $t'/t$,
as appropriate for cuprates, favors 
partially filled (metallic) stripes for both vertical and diagonal 
configurations. 
At around optimal doping diagonal stripes, 
site centered (SC) and bond centered (BC) vertical stripes become degenerate 
suggesting strong lateral and orientational fluctuations.
We find that within the GA
the resulting phase diagram is in agreement with experiment whereas it is not
in  the Hartree-Fock approximation due to a 
strong overestimation of the stripe filling.
Results are in agreement with previous calculations within 
the three-band Hubbard 
model but with the role of SC and BC stripes interchanged.  
\end{abstract}
\maketitle
\section{Introduction}
The existence of stripes, i.e. antiphase domain walls in an antiferromagnet 
stabilized by doped holes, is now a well established fact in
cuprate\cite{tra95,tra96,tra97,yam98,ara99,ara00} and 
nickelate\cite{nio1,nio2,nio3} materials. 

Direct evidence for static stripes in cuprates 
has been more clearly established in La$_{2-x}$Sr$_x$CuO$_4$
(LSCO) eventually co-doped with Nd or Eu.
However incommensurate inelastic neutron scattering signals 
(widely interpreted as evidence for dynamical stripe fluctuations) 
have been found in both LSCO and YBa$_2$Cu$_3$O$_{7-\delta}$ (YBCO) 
compounds with a remarkably similar 
phenomenology.~\cite{dai01} Even more static 
charge order has been found in YBCO without the need for additions 
like Nd.~\cite{moo02}

In LSCO the inverse stripe spacing grows 
linearly with doping $x$ up to $x\approx 1/8$
and the orientation rotates from vertical (i.e. oriented along the
Cu-O bonds) by $45^0$ to diagonal for concentrations lower 
$x \approx 0.05$.~\cite{yam98,mat00,fuj02} 
Moreover, from this linear relation it turns out
that in the LSCO compounds  stripes are characterized by one doped
hole per two unit cells along the domain wall, that is a linear density 
of added holes $\nu  \approx 0.5$ (hereafter called 'half-filled stripe').

Regarding the Bi$_2$Sr$_2$CaCu$_2$O$_8$ (BiSCCO) family it is a very 
difficult material to perform neutron scattering so 
few experiments exist [see the discussion in  Ref.~\onlinecite{he01}].
Interestingly this study finds a peak much broader than the momentum 
resolution  leaving plenty of room for incommensurate effects.
The existence of incommensurate scattering in BiSCCO is also supported by
the experiments of Ref. \onlinecite{mo00}.
While most neutron scattering experiments focus on the structure of 
spin excitations, the inhomogeneous charge distribution as arising 
from the formation of 
stripes has also been detected by local probes like 
NQR \cite{kraem99,sing02} and NMR.~\cite{has02} 
Due to refinements in the experimental technique Haase et al. \cite{has02} 
where even able to demonstrate a correlation of charge and
density variations on short length scales.

All these experiments suggest that stripes are a common phenomena of all
cuprates families and therefore may be related to superconductivity. 

Concerning the theoretical aspects, stripes were predicted by the
inhomogeneous Hartree-Fock (HF) approximation  in the 
three-band Hubbard model\cite{zaa89}, the one-band Hubbard\cite{mac89,hsch90}
and the $t-J$ model.~\cite{pol89} However these pioneering studies predicted
insulating stripes with a linear density  of $\nu=1$ added holes per lattice 
constant along the stripe (instead of $\nu\sim 0.5$)
which led to an early rejection of stripes.~\cite{swche91} 

Recently we have presented a computation of metallic mean-field
stripes\cite{lor02b} within an unrestricted Gutzwiller
approximation\cite{gut65} (GA) applied to the three-band Hubbard model.
The behavior of the magnetic incommensurability 
$\epsilon\equiv 1/(2d)$\cite{tra97,yam98,ara99,ara00}
($d$ is the distance between charged stripes in units of the
lattice constant), chemical
potential\cite{ino97,har01}, and transport
experiments\cite{nod99,wan01} as a function of doping has been
explained in a parameter free way.~\cite{lor02b} Addition of
RPA-type fluctuations within a recently developed 
time dependent GA\cite{sei01} 
has explained\cite{lor03} also the evolution of the optical 
conductivity with doping\cite{uch91,suz89} in a broad frequency range. 

In our previous computations\cite{lor02b,lor03}
we fixed the 
parameter set  of the three band Hubbard model to the values derived
for the LSCO family within the local density approximation (LDA).~\cite{mcm90} 
In this sense we view our previous work as the last stage of an 
{\it ab inito} determination of the electronic structure. 
On the contrary we would like to explore in the present paper 
the parameter dependence of the results. 
This kind of investigation is interesting not only in order to
check  the robustness of the results but also to obtain  hints about the 
expected trends in the behavior among the different cuprates materials. 
The complexity of the three band model, however, makes it 
difficult to perform such study.
To achieve our goals we adopt here a much simpler Hamiltonian,
namely the extended one-band Hubbard model. Given the popularity
of the model it is also interesting to explore, to what 
extent the physics found in the three band Hubbard model
can be found in the one band counterpart. In the present
work we find that  for realistic parameters a
similar phase diagram is found in the one-band and the three-band
models\cite{lor02b} however some important details about the 
symmetry of stripes differ.

Regarding the material dependence it has been proposed 
recently\cite{pav01} that a one-band model
description of the various cuprate families essentially 
differs in the ratio between next-nearest and nearest neighbor hopping
$t'/t$. In fact it is found that the transition temperature scales
with $t'/t$ ranging from $t'/t \approx -0.15$ for the LSCO single layer
compound up to $t'/t \approx -0.4$ for the Tl and Hg based materials. 
We show that this parameter plays a key role for the stability and electronic
structure of the stripes suggesting a possible connection between stripes and 
superconductivity. 
In particular the higher the $t'$ the smaller
is the optimum filling $\nu$ of the stripe (Fig.~\ref{nuopt}).  
Additionally we present a thorough comparison
of the stability of stripe and polaron textures upon varying $t'/t$.

The influence of a next-nearest neighbor hopping term on the 
stripe formation has been previously investigated by dynamical mean-field 
theory (DMFT)\cite{fle01}
and the HF approximation.~\cite{ichi99,val00,norm02}
It was found \cite{fle01b,norm02} that increasing the ratio
$|t'/t|$ leads to a suppression  of static stripes. This may indicate
a more dynamical character of stripes in some systems like 
Tl and Hg based compounds. 

In case of the HF approximation\cite{inu91} 
vertical stripe solutions are only favored for unrealistic small
values of $U/t\approx 3 ... 5$ whereas a ratio of $U/t\approx 10$
is required in order to reproduce the low energy spectrum of the
three-band model.~\cite{hyb92}
Therefore our investigations here are based on the unrestricted Gutzwiller
approximation \cite{sei98} which provides an excellent variational 
Ansatz for the ground state energy of the Hubbard model. We compare
our results with the HF approximation an show that 
HF cannot account for the site centered stripe
topology as relevant in the one-band model.

In Sec.~\ref{model} we briefly outline our method and present the results
of our calculation in Sec.~\ref{results}. Finally we conclude the discussion
in Sec.~\ref{conc}.

\section{Model and Formalism}
\label{model}
We consider the two-dimensional Hubbard model on a square lattice, with
hopping restricted to nearest ($\sim t$) and next nearest ($\sim t'$) 
neighbors 
\begin{eqnarray}\label{HM}
H&=&-t\sum_{<ij>,\sigma}c_{i,\sigma}^{\dagger}c_{j,\sigma}\\
&-&t'\sum_{<<ij>>,\sigma}c_{i,\sigma}^{\dagger}c_{j,\sigma} 
+ U\sum_{i}
n_{i,\uparrow}n_{i,\downarrow}.\nonumber
\end{eqnarray}
Here $c_{i,\sigma}^{(\dagger)}$ destroys (creates) an electron
with spin $\sigma$ at site
$i$, and $n_{i,\sigma}=c_{i,\sigma}^{\dagger}c_{i,\sigma}$. $U$ is the
on-site Hubbard repulsion which in the following is fixed to the
value $U/t=10$ as relevant for the cuprates (see e.g. Refs. 
\onlinecite{hyb92,dag94}).  
The unrestricted Gutzwiller approximation (GA) can be implemented by
either a variational Ansatz \cite{geb90} or the Kotliar-Ruckenstein
slave-boson scheme \cite{kot86} and results in the following
energy functional
\begin{eqnarray}\label{EGW}
E^{GA}(\rho,D)&=&-t\sum_{<ij>,\sigma}
z^{GA}_{i,\sigma}z^{GA}_{j,\sigma} \rho_{ij,\sigma}\\
&-&t'\sum_{<<ij>>,\sigma}
z^{GA}_{i,\sigma}z^{GA}_{j,\sigma} \rho_{ij,\sigma}
+U\sum_{i}D_{i} \nonumber
\end{eqnarray}
where $\rho_{ij,\sigma}=\langle c_{i,\sigma}^{\dagger}c_{j,\sigma}\rangle$
denotes the single-particle density matrix, $D_i$ are variational
parameters related to the double occupancy of the sites and
the Gutzwiller hopping factors are given by
\begin{equation}
z^{GA}_{i,\sigma}=\frac{\sqrt{(1-\rho_{ii}+D_i)(\rho_{ii,\sigma}-D_i)}
+\sqrt{D_i(\rho_{ii,-\sigma}-D_i)}}{\sqrt{
\rho_{ii,\sigma}(1-\rho_{ii,\sigma})}}.
\end{equation}
The energy functional Eq. (\ref{EGW}) has to be minimized with respect to
the double occupancy parameters $D$ and
the density matrix $\rho$ where the latter
variation has to be constrained to the subspace of Slater
determinants. For technical aspects of this variational
procedure we refer to Ref. \onlinecite{sei98}.

The clusters investigated in the following have dimensions
$N=N_x\times L$ up to $16\times 16$ and we use periodic
and antiperiodic boundary conditions in order to stabilize
homogeneous (metallic) stripe textures.

\section{Results}
\label{results}
\subsection{Charge and spin profile of stripes}
\begin{figure}[tbp]
\includegraphics[width=8cm,clip=true]{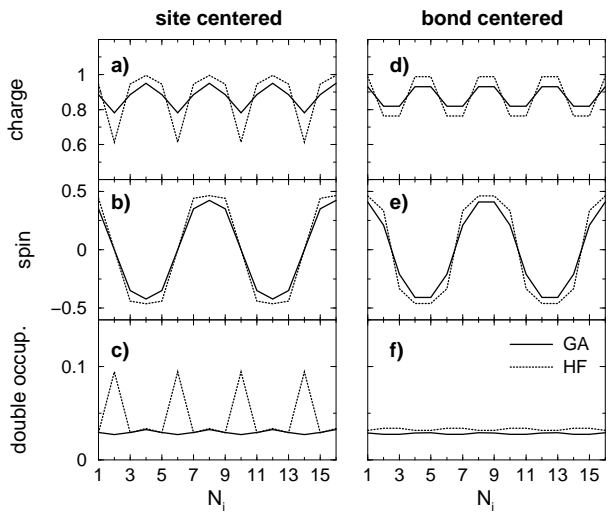}
\caption{Profile of the charge density (a,d), 
staggered spin order parameter (b,e), and
double occupancy (c,f) of site centered (a-c) and bond centered (d-f) stripes.
Results are for $t'/t=-0.2$ and $n_s=4$ half-filled stripes on a 
$16\times 16$ lattice  corresponding to $d=4$ and $x=1/8$.}
\label{rdx}
\end{figure}

In Fig.~\ref{rdx} we report the typical charge- and spin profiles in a 
cut perpendicular
to the stripe for vertical site (SC) and bond (BC) centered 
stripes as computed 
with HF and GA respectively.
All vertical stripe solutions investigated in this paper are 
metallic (i.e. homogeneous along the stripe) except for integer $\nu$ and we
have suppressed 1-D instabilities by choosing appropriate
boundary conditions (i.e. periodic or antiperiodic in direction
of the stripes).

The total charge shows dips at the central site (sites) of the
antiphase SC (BC) domain walls (Fig.~\ref{rdx}a,d) corresponding to the 
accumulation of holes.   
As also found in previous studies
\cite{sei98,lor02b,sad00} the GA charge modulation is 
much weaker as compared to HF. 

Due to symmetry of the antiphase domain walls
the spin density is zero along the stripe 
in case of SC stripes (Fig.~\ref{rdx}a-c). Therefore 
a large charge reduction is the
only possibility to reduce the HF double occupancy on these
sites. On the other hand the double occupancy within the GA 
can be separately tuned  by the variational parameter $D$ leading to
a smaller charge modulation amplitude  with respect to HF. 
In Fig.~\ref{rdx}c we show the smaller double 
occupancy at the core stripes sites (like site 2) for the GA 
whereas both HF and GA acquire approximately the same value within 
the AF ordered regions. 
In contrast, the BC stripe solution has finite spin polarization on 
all sites (Fig.~\ref{rdx}e) so that in this case 
a large HF double occupancy is avoided (Fig.~\ref{rdx}f). 
This has important consequences for the respective stability of
BC and SC stripes within the two approximations as we will see below.

\subsection{Effect of $t'$ on the band structure}

In order to understand the role of $t'$ in the
context of stripe formation we proceed by investigating
the influence of this parameter on the electronic structure.
Within our variational (mean-field) approach the undoped
system is characterized by a homogeneous spin-density
wave state which band structure is characterized by
a lower (filled)  and an upper (empty) Hubbard band,
separated by a single particle energy gap $\Delta$. Due to the incorporation
of correlation effects already on the mean-field level, 
$\Delta^{GA}$ is significantly smaller than
within the HF approximation where it is simply given by
$\Delta^{HF}=U m$ and $m$ denotes the magnetic moment.

\begin{figure}[tbp]
\includegraphics[width=8cm,clip=true]{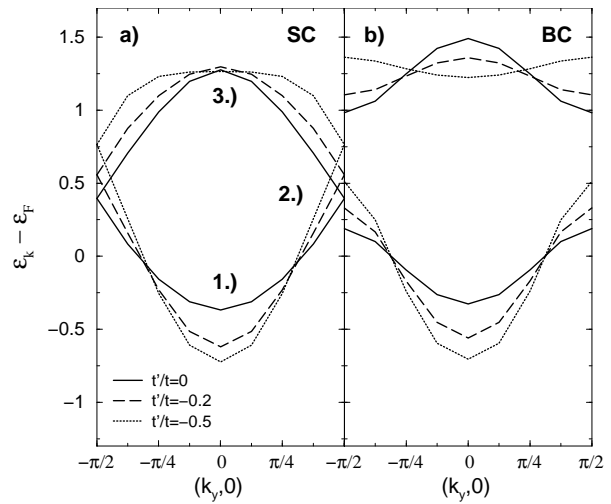}
\caption{Dispersion of the mid-gap bands along vertical
$d=4$ stripes in a $16\times 16$ lattice computed within the GA. 
$t'/t=0,-0.2,-0.5$.
a) Site centered stripes; b) Bond centered stripes}
\label{bands}
\end{figure}

The stripe state at finite doping induces the formation of 
two bands within the Mott-Hubbard gap. These are almost
flat perpendicular to the stripes but some dispersion
develops when the stripes become closer, similar to
what is found in the 3-band model.~\cite{lor02b}
In Fig.\ref{bands} we show the dispersion for both bands in the
stripe direction ($y$) for various values of $t'/t$.

\begin{figure}[tbp]
\includegraphics[width=7cm,clip=true]{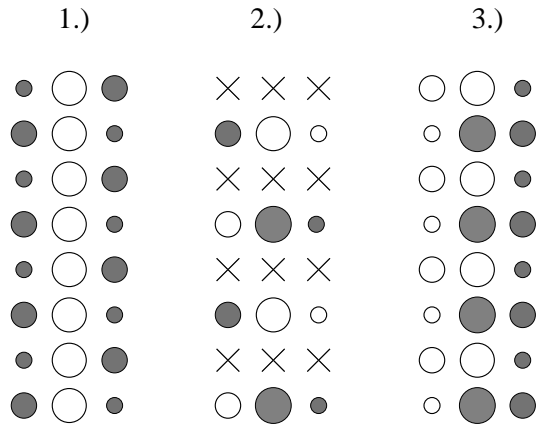}
\caption{Sketch of the amplitudes $\Phi_i(k)$ along site
centered stripe
sites $i$ for selected k-states as indicated in the band structure
of Fig. \protect\ref{bands}a. 1.)$\equiv k=(0,0)$ of the active band;
2.)$\equiv k=(\pi/2,0)$; 3.)$\equiv k=(0,0)$ of the upper band.
Empty (full) circles correspond to positive (negative) amplitudes
which value is indicated by the radius. Crosses mark $\Phi_i(k)=0$.}
\label{amp}
\end{figure}
Doped holes occupy states in the lower band (hereafter called the
active band) and a half-filled stripe corresponds to a
half-filled active band.
Furtheron the bandstructure of the BC solution (Fig. \ref{bands}b) 
displays a gap between 
both bands which is due to a AF spin-density modulation along the
BC stripe.
For both, SC and BC stripes, Fig. \ref{bands} reveals a significant broadening
of the active band upon increasing $|t'|$.
This broadening has a 
strong influence on the kinetic energy of holes in the stripe and 
therefore also on the density of states at the Fermi energy and, as we
will show below, on the value of the optimum $\nu$.

The broadening can be understood by looking at the single particle 
wave functions (WF) of selected k-points of the active band as
shown in Fig. \ref{amp} for the case of SC stripes. 
The $k=0$  WF is characterized 
by having opposite sign at the core of the stripe than on the adjacent 
(AF ordered) legs. 
As a consequence a positive next-nearest neighbor
hopping (i.e. negative $t'$, cf.  Eq. (\ref{HM})) leads to a lowering 
of this state in first order perturbation theory.
By the same argument one can show that the zone boundary state of the
active band is not affected by $t'$ since next-nearest neighbor
hopping along the stripe always connects to a site $i$ with a node in
the single particle wave function
(the shift in energy in Fig. \ref{bands} 
is due to the fact that $E_F$ shifts downwards with $t'$).
For a sizeable SDW gap $\Delta$ on the adjacent legs of the stripe 
the broadening of the active band can be evaluated
as $\delta B = 36t' t^2/\Delta^2$. Since $\Delta^{HF} > \Delta^{GA}$
as already mentioned above, the $t'$ band width renormalization is
expected to be larger in the GA as compared to HF.

For completeness we mention that  
similar considerations also hold for the upper band in the
Mott-Hubbard gap. From the structure of the $k=(0,0)$ wave-function  
(cf. Fig. \ref{bands}) it turns out that a negative
$t'$ induces a shift of this state to lower energies (in Fig. \ref{bands}
this shift is almost compensated by the change in the chemical potential).
Correspondingly we find a narrowing of the upper band 
$\delta B = -28t' t^2/\Delta^2$ which for large $t'$ leads to the
flat structure as observed in Fig. \ref{bands}.

\subsection{Parametrization of $e_h$; GA vs. HF}

To determine the more stable solution among different 
textures we need to compare the total energies per atom, $E/N$,
for the same doping, $x\equiv N_h/N$. 
Equivalently we can compare the excess energy per hole with
respect to the undoped  antiferromagnet (AF), hereafter
the ``energy per hole'':
\begin{equation}  
e_h=\frac{E(N_h)-E_{\rm AF}}{N_h}
\end{equation}  
where $E(N_h)$ is the total energy of the inhomogeneous 
state (generally a stripe state) with $N_h$ holes 
counted from half-filling and  $E_{\rm AF}$ is the AF 
solution on the same cluster.
The minimum of $e_h$  at low doping
(non-interacting stripes) has the
following meaning: it determines the optimum filling that stripes 
should have in order to accommodate 
a given number of holes in a variable number of stripes.

\begin{figure}[htbp]
\includegraphics[width=8cm,clip=true]{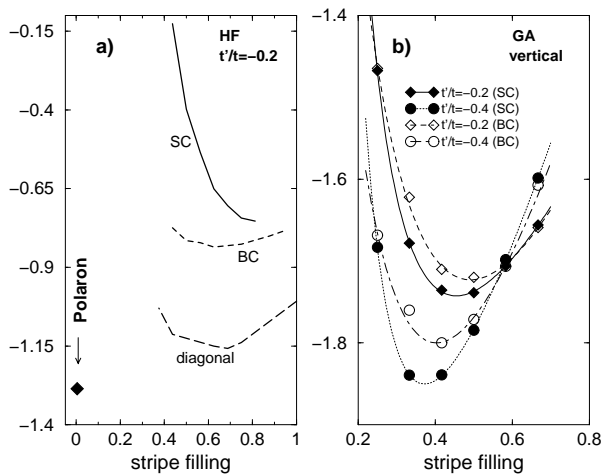}
\caption{(a) HF energy per hole $e_h$ for vertical stripes 
as a function of the
stripe filling $\nu$ and for various values of $t'/t$. 
Results in (a) are for 2 $d=5$ stripes on a $10\times 10$ lattice.
(b) GA energy per hole for 4 site centered (SC) and bond
centered (BC) stripes on a $12\times 12$ lattice (symbols).
The lines correspond to fits obtained with Eq. (\protect\ref{ehnu})}
\label{ehhf}
\end{figure}

In Fig.~\ref{ehhf} we report the  energies per hole for
different configurations in the HF and in the GA approximations as a
function of the stripe
filling $\nu=N_h/(n_s L)$. Here  $L$ is the linear dimension 
in the $y$ direction of the cluster which coincides with 
the stripe length, $n_s$ denotes the number of stripes
and $N_h$ is the total number of holes (counted from half-filling) 
in the system.  
More detailed results for vertical SC
and diagonal stripes in the  
 GA  are reported in Fig.~\ref{d35fig}.

In order to analyze these results we expand the excess energy 
per unit length as a Taylor series in $\nu$:
\begin{equation}\label{enu}
\frac{E(N_h)-E_{\rm AF}}L=A+B \nu + C \nu^2 .
\end{equation}
The energy per hole in a stripe thus follows the relation
\begin{equation}\label{ehnu}  
e_h=A/\nu + B + C\nu
\end{equation}  
which provides an excellent fit to the data points
as shown in Figs. \ref{ehhf}b, \ref{d35fig}.
The minimum of $e_h$ is given by 
$$\nu_{min}=\sqrt{A/C}$$
and the second derivative of the energy per hole at the minimum
reads as 
\begin{equation}\label{e2dnu}
e''_{h}(\nu_{min})=2C\sqrt{C/A}.
\end{equation}  
Note, however, that the single particle spectrum has a gap for $\nu=1$.
Therefore the energy has a cusp-like singularity at this filling
which causes the $\nu=1$ state to be the minimum for a wide 
range of parameters. We can take this into account by modifying the
above formula to: 
\begin{equation}\label{numin}  
\nu_{min}=\min(\sqrt{A/C},1)
\end{equation} 
In the dilute limit, corresponding to large stripe separation, 
the coefficients can be interpreted as follows:
$A$ is the energy cost per unit length 
to create an antiphase domain wall in the
{\it stoichiometric} (otherwise AF ordered) system.
The parameter $B$ can be defined as the chemical potential to add 
one hole into the ``empty'' stripe. 
Finally the parameter $C=C_K+C_I$ is related to the kinetic ($C_K$) and 
interaction ($C_I$) energy of the system.

In order to separate the two contributions $C_K$, $C_I$ to the
parameter $C$ we 
have adopted the following procedure.
For a small value of $\nu$ we have optimized the variational 
parameters of the GA. Then  the energy
as a function of $\nu$ has been computed {\it without} 
changing the variational 
parameters, i.e. by solely filling the frozen bands. The resulting 
quadratic contribution to the energy gives the $C_K$ 
parameter and since $C$  is known we also obtain $C_I$.
 Within mean-field the 
latter is due to readjustments of the charges and double occupancies
in response to the added holes.

For the parameters under consideration we 
find that $C_I$ is approximately $2-3$ times larger than $C_K$.
However, it is important to note that $C_I$ is almost independent
from the next-nearest neighbor hopping, whereas from Fig.
\ref{bands} it is obvious that the kinetic energy of the 
holes moving along the stripe will have a significant   
dependence on $t'$. To see more explicitly how the active 
bandwith reflects on the kinetic energy one can 
approximate the active band density of states per 
spin by a rectangular shape of height $1/W(t')$. Here  $W(t')$ 
represents an effective bandwith which is roughly given by the width
of the lower (active) band in Fig.~\ref{bands}. One finds
that the kinetic energy is quadratic in $\nu$ and 
$C_K \sim W(t')$. Below we discuss in detail how $t'$ determines the 
stability of stripes via this mechanism. 

To estimate $A$ we need an estimate of the energy to create
an empty domain wall. Since this is related to an excitation of the 
undoped system we can use the Heisenberg model with a magnetic
interaction $J=4t^2/U$ in the limit of $U>>t$. The estimate is easily 
done in the case of 
BC stripes which have a core with  ferromagnetic alignment 
along the domain wall. In the Ne\'el limit one obtains: 
 $A=J/2$.

The case of SC stripes is more subtle. Indeed for SC stripes 
one has a core without permanent moments
which, contrary to the BC case, does not have  a classical analog
within a  Heisenberg description of the undoped state. A non classical 
state will consist of a spin liquid along the core which
in any case will give $A\sim J$. Within the HF approximation
of the Hubbard model  
the corresponding solution has a non-magnetic core
with maximum double occupancy (1/4). This is at odds with the
Heisenberg picture of a spin liquid core which, by construction, implies
small double occupancy for the original fermions.  Therefore the 
HF approximation introduces an error in $A$ which in the large $U$
limit is of the order of $U/4$. This is due to the fact that moment
formation is the only mechanism within HF to reduce double occupancy. 
On the other hand the GA allows an unpolarized state with small 
double occupancy ($D\sim t^2/U^2$) thanks to the  additional 
variational parameters $D_i$ as shown in Fig.~\ref{rdx}c,f.
Due to this mechanism we obtain $A$ within the GA of the correct
order $A\sim U D\sim J$. We see in Fig.~\ref{ehhf} that the 
 curves for SC and BC stripes are similar in the GA whereas in HF 
metallic SC stripes are energetically unfavorable with respect to
BC (and diagonal) ones. 
Clearly the latter faultiness of the HF approach is due to 
the overestimation of the parameter $A$ mentioned above 
[c.f. Eq.~\ref{numin}] which leads to an unphysical large value of
 $\nu_{min}$. For this reason we conclude that HF theory
is not an adequate approach for the description of SC stripes. 

Within the HF approximation it is known \cite{inu91}
that for $U/t >3.6$ vertical stripes are less stable than 
diagonal ones. Moreover spin polarons become the ground state when 
$U/t > 8$. Fig. \ref{ehhf}a demonstrates that similar results 
hold when one includes next-nearest neighbor hopping. As we will show
below the situation is quite different in the GA approximation 
which also allows a comparison between SC and BC stripes on an equal
footing. 

It is seen from Fig. \ref{ehhf}b that within the
Gutzwiller approach  SC stripes become degenerate with BC 
for large {\it and} small filling factors  $\nu$.
Especially it turns out that  BC empty stripes are slightly lower in energy 
than SC empty stripes. However, SC stripes tend to have larger 
bandwith than BC ones (cf. Fig. \ref{bands}) which leads to a larger $C$, 
larger curvatures of $e_h$ vs. $\nu$
[Eq .~(\ref{e2dnu})] and hence smaller values of $\nu$. The difference in  
curvature allows for the two intersections between the $e_h$ curves
for BC and SC solutions in Fig. \ref{ehhf}b.
We will show below that the crossing at large $\nu$ corresponds to
optimal doping and that in the underdoped case SC stripes are more
stable than BC ones. 
The stability of SC solutions with respect to BC ones at low doping 
as well as the increased degeneracy for larger concentration
is in agreement with DMFT calculations in the one-band
model.~\cite{fle00} Therefore it seems to be a robust feature 
of the model and not to depend on the approximation.  
On the other hand it is opposite to what is observed
in the 3-band model\cite{lor02b} where BC stripes
constitute the most stable low doping structure.

\subsection{Effect of $t'$ on the stability of stripes}

\begin{figure}[tbp]
\includegraphics[width=7cm,clip=true]{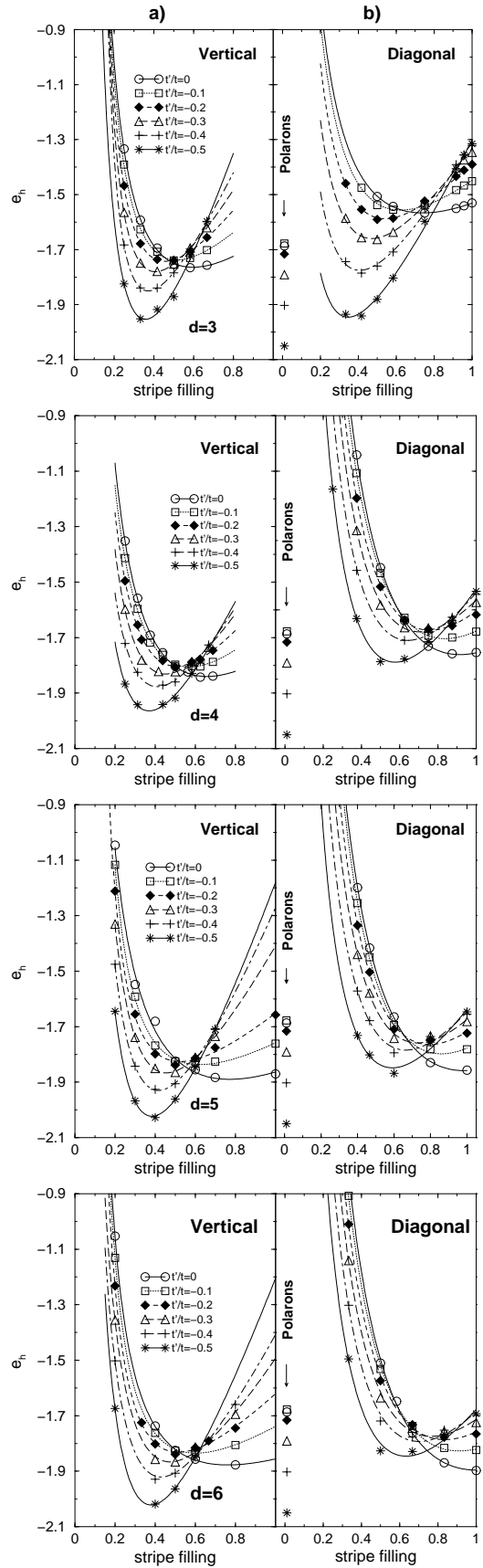}
\caption{GA excess energies per hole $e_h$ for vertical SC (a) and
diagonal (b) stripes as a function of the
stripe filling $\nu$ and for various values of $t'/t$. 
The stripe separation (in lattice constants) is
$d=3,4,5,6$ (from top to bottom).  We also report the energy
of the polaron solutions in the right panel.}
\label{d35fig}
\end{figure}

We now proceed to discuss in detail the role of $t'$ on $e_h$.  
Due to the dependence of $C_K$ on $t'$ and the constancy of the parameter
$A$ 
we observe in Fig. \ref{d35fig} a shift of optimal
stripe filling to smaller values  and a steeper minimum
upon increasing $|t'|$. Hence, the larger is $|t'|$ the more stable are
partially filled stripes with respect to their filled (insulating) 
counterpart in
agreement with the estimate $C_K\sim W\sim |t'|$ and Eq.~(\ref{numin}).
One can understand the  reason for this behavior by comparing the 
two cases $\nu=1$ (insulating) and $\nu=1/2$ (metallic), remembering
that  the former structure has twice as much holes per stripe.
For the sake
of simplicity let us neglect for the moment the interaction part $C_I$. 
Then, if the cost to create a 
stripe dominates ($A>C$ or roughly speaking $J>W$), 
$\nu=1$ stripes are more convenient because this minimizes the
number of stripes that have to be created to accommodate 
a given number of holes.  
On the other hand for $\nu=1$ the Pauli exclusion principle 
forces to fill the higher kinetic energy states of the active band,
which for broad bands  becomes energetically unfavorable. 
As a consequence it is more convenient for systems with broad bands
($W>J$)  to pay the energy $J$ in order to create more 
stripes, thus reducing the filling per stripe and avoiding the occupancy 
of high kinetic energy states of the active band.  
Clearly the relevant parameter which controls this behavior 
is $J/|t'|\sim J/W \sim A/C$ with small ratio  $A/C$ favoring 
partially filled stripes in agreement with Eq.~(\ref{numin}) (see also 
Fig.~\ref{nuopt} below). We note that this is in agreement with our 
findings in the 
three-band Hubbard model where the role of $t'$ was played by the
oxygen-oxygen hopping.~\cite{lor02b}

The energy of (vertical) metallic stripes has to be compared with other
saddle-points of the GA energy functional.
It should be noted that especially at large doping
(but depending also on $t'/t$) there exists a huge variety of 
inhomogeneous charge
and spin textures which are rather close in energy and compete
with the vertical stripe solution.
Among this variety we have selected diagonal stripes and spin polarons
which energy per hole is reported in Fig. \ref{d35fig}b.
This choice is motivated by the idea that with increasing kinetic
energy the stripes should undergoe some kind of melting process 
which reflects in substantial lateral
and orientiational fluctuations and an eventual decay
into individual segments.
However, in our static approach we can only access some 'snapshot' 
configurations 
of this melting process and demonstrate the increased degeneracy in energy 
of these structures with doping.

Two major points result from Fig. \ref{d35fig}: {\it i}) For fixed stripe
periodicity $d$ vertical solutions are in  general lower in energy
than diagonal ones.~\cite{note} 
However, for larger stripe separation ($d=5,6$)
and small ratio $|t'/t|\lesssim 0.1$ one observes from Fig. \ref{d35fig} that 
diagonal and vertical structures are rather close in energy.
{\it ii}) For moderate 
values of $|t'/t|\lesssim 0.4$ spin polarons are always higher 
in energy than (vertical) stripe structures
but the polaron solution becomes more stable 
for larger $|t'/t|$ depending
on the stripe separation. At low doping where the stripes are
almost noninteracting ($d >5$) the critical value for
(vertical) stripe stability is  $t'/t\approx -0.4$.
In this region of parameters  the ground state will have
contributions from both stripes and dissociated holes. We can speculate that
in this case, going beyond mean-field,
stripes will have a more dynamical character. In this sense
our finding is in agreement with previous investigations 
\cite{fle01b,norm02} which found a suppression
of static stripe formation upon increasing $|t'/t|$. 
Note that for finite doping the polaronic energy 
naturally depends on the special arrangement of the 
self-trapped charge carriers on the lattice. However, for
the selected large value of $U/t=10$ the spin-polarons are quite
localized and therefore their energy  is almost independent of
doping up to hole concentrations $x\approx 0.2$.

Fig.~\ref{nuopt} displays the optimal filling $\nu_{opt}$ 
(corresponding to the curve minima in Fig. \ref{d35fig})
versus the stripe separation for different next-nearest neighbor hoppings.
\begin{figure}[tbp]
\includegraphics[width=8cm,clip=true]{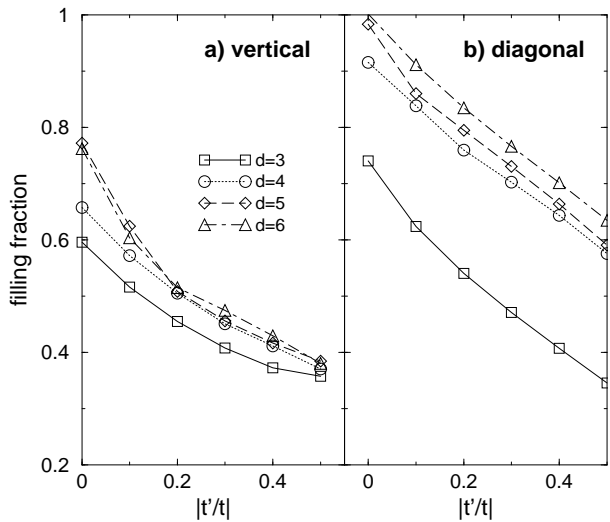}
\caption{Optimal stripe filling as a function of periodicity
for (a) vertical and (b) diagonal stripes.}
\label{nuopt}
\end{figure}
For stripe distances $d>4$ the vertical stripes are
almost non-interacting and therefore the filling fraction
versus $t'/t$ is the same as can be seen from Fig.~\ref{nuopt}a.
For smaller distances stripes start to repel,
resulting in a shift of $e_h$ to higher values (cf. Fig. \ref{d35fig}a).
However, the increased inter-stripe interaction leads also to a 
larger dispersion perpendicular to the stripes 
which in the $e_h$ curves of Fig. \ref{d35fig} 
reflects as an increase in curvature
(i.e. in the 'kinetic' constant C) upon going from
$d=4$ to $d=3$ stripes. As a result one therefore observes 
a shift of $\nu_{opt}$ to lower values for the $d=3$ solution.

The situation for diagonal stripes is qualitatively similar,
i.e. upon increasing $|t'/t|$ the minimum of $e_h$ shifts to 
lower values of $\nu$. Since the kinetic energy in diagonal direction
(and thus the parameter $C_{diag}$) is naturally smaller than along the bonds 
we find the optimal filling at larger values as compared to vertical
stripes. 
Moreover, diagonal stripes are more extended so that
a sizeable inter-stripe interaction is present even for 
periodicities $d=5,6$.
In Fig. \ref{d35fig}b this reflects as a 
continuous decrease of the $e_h$ curves upon increasing $d$. 
For the same reason $\nu_{opt}$ continuously increases with $d$ for
fixed $t'/t$ as can be seen from Fig. \ref{nuopt}b.

\begin{figure}[tbp]
\includegraphics[width=7cm,clip=true]{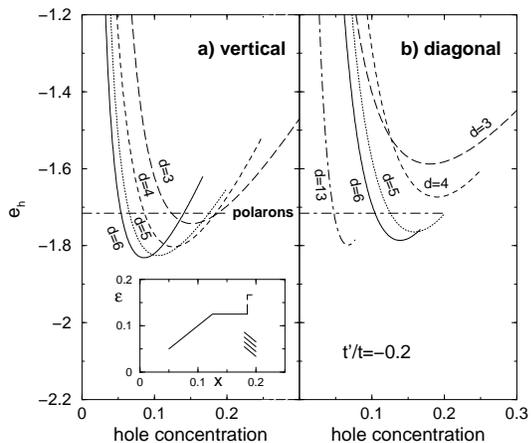}
\caption{Energies per hole for vertical (a) and diagonal (b)
stripes as a function of doping. $t'/t=-0.2$ as 
appropriate for LSCO.
The inset displays the incommensurability versus doping
as deduced from (a). The diagonal pattern indicates the degeneracy
with diagonal stripes.}
\label{finfig}
\end{figure}

Comparing Fig. \ref{nuopt}a with the experimental situation in
the LSCO high-T$_c$ compound it thus turns out that the phenomena 
of half-filled
vertical stripes for $d \ge 4$ requires for the ratio
between next-nearest and nearest neighbor hopping $t'/t \approx -0.2$. 
For this value Fig. \ref{finfig} reports the corresponding $e_h$ curves 
but now plotted as a function of the hole doping.
Since the charge modulation of vertical stripes has 
a width of about 4-5 lattice constants
the inter-stripe interaction is vanishingly small  for $d \ge 5$ and thus the 
energy depends only weakly on distance. 
Moreover, from Fig. \ref{d35fig} we have seen that in this regime 
the ground state solution
has always filling close to $\nu=1/2$ and thus the incommensurability 
\begin{equation}
  \label{eq:edx}
  \epsilon=\frac{x}{2\nu}.
\end{equation}
behaves as $\epsilon\approx x$ in agreement 
with experiment.~\cite{tra97,yam98,ara99,ara00}
For $d \le 4$ ($x \ge 1/8$)
stripes overlap (see Fig.~\ref{rdx}) which leads to a vertical 
shift of the curves as shown in Fig.~\ref{finfig}a.
Hence it is unfavorable for the system
to increase the density of stripes from $d=4$ to $d=3$, the incommensurability 
gets locked at $\epsilon=1/8$ ($d=4$) and the filling $\nu$ of the stripes 
starts to increase. 
This feature favors the degeneracy between bond and
site centered vertical stripes as can be deduced from Fig. \ref{ehhf}b
suggesting the importance of lateral stripe fluctuations in this doping range.
Disregarding for the moment the polaron solution we observe a 
further transition toward $d=3$ stripes at $x \sim 0.2$. 
A similar result has been found also in the 3-band model.~\cite{lor02b}

Diagonal stripes are more extended so that
the corresponding $e_h$ curves start to shift for $d \le 5$ already.
In the low doping regime they are less stable than
vertical stripes, however, the energy difference becomes
rather small for $x \gtrsim 0.2$. 
Moreover, both vertical and diagonal stripe solutions approach the
energy of the polaron solution at this particular doping. 
This means that for concentrations $x>0.2$ the energy is
almost independent of the charge topology. The GA landscape 
is not dominated by a well defined saddle point and an associated Slater 
determinant but a  large number of energy minima with very different 
charge topologies compete. 
In this regime the ground state will thus have contributions 
from all Slater determinants
corresponding to the different charge configurations with equivalent
energy. We expect a state which has lost the orientational 
order information.~\cite{kiv98} 

In the low doping regime neutron scattering experiments
on LSCO observe a change of the incommensurability vector below
$x \approx 0.05$ from the vertical to the diagonal
direction.~\cite{yam98,mat00,fuj02} 
From the results as shown in Fig. \ref{finfig} we 
cannot infer a crossover towards diagonal solutions at low doping. 
However, the closeness in energy of vertical and diagonal structures 
suggests that the orthorombic distortion, which continuously
increases upon underdoping in LSCO may lead to a stabilization
of the diagonal stripe phase at small hole concentrations
(see Normand et al. \cite{norm01} for a detailed
discussion of the relation between stripe orientation and
lattice distortion).
Moreover, we believe that other effects, not included in our
model,  are important in this regard. 
Indeed the contribution of long-range Coulomb interactions
is expected to become important at low doping. A point charge estimate
shows that this 
tends to favor diagonal configurations over
vertical ones. 

To summarize the results for $t'/t=-0.2$:  We obtain 
half-filled vertical stripes and $\epsilon=x$
up to hole concentrations $x=1/8$ and increasing $\nu$ and 
$\epsilon=1/8$ for 
$1/8<x<0.2$ in agreement with experiment\cite{tra97,yam98,ara99,ara00}
and with our previous 3-band computations.~\cite{lor02b}
The behavior  for the filling implies that the chemical potential $\mu$
will be approximately constant for $x<1/8$ and decrease with doping for
$x>1/8$ which is also supported by experiment.
\cite{ino97,har01} However, it should be noted that the theoretically
obtained shift in $\mu$ exceeds the experimental one by a factor of $2 ...3$
which may be attributed to finite size effects. A similar discrepancy 
is also observed in the 3-band model and we refer to
Ref. \onlinecite{lor02b} for a further discussion of this point.

\begin{figure}[tbp]
\includegraphics[width=7cm,clip=true]{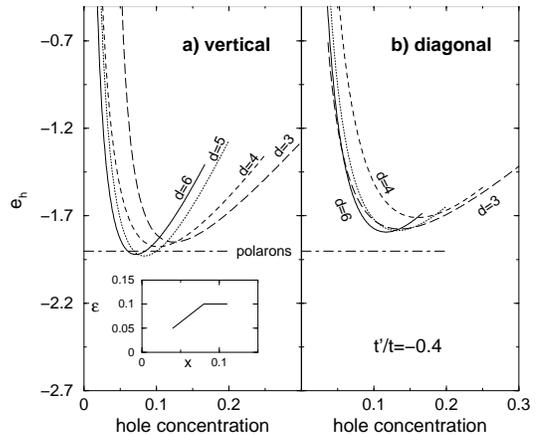}
\caption{Energies per hole for vertical (a) and diagonal (b)
stripes as a function of doping. $t'/t=-0.4$ as appropriate
for Bi2212.
The inset displays the incommensurability versus doping
as deduced from (a).}
\label{finfig2}
\end{figure}

Finally, in Fig. \ref{finfig2} we show the phase diagram calculated for
$t'/t=-0.4$ which is considered to be the appropriate ratio
for the Tl$_2$Ba$_2$Ca$_2$Cu$_3$O$_10$ and
HgBa$_2$CaCu$_2$O$_6$ compounds.~\cite{pav01}
In this case low doping vertical stripes have a filling fraction of
$\nu \approx 0.4$ so that the incommensurability $\epsilon \approx 1.25 x$
is expected to grow faster than for the LSCO compounds.
However, already the $d=4$ structures are unstable with respect to
the spin polaron solution and thus the incommensurability starts to
saturate for $x\ge 0.08$ corresponding to the $d=5$ structure.
Moreover since the polaron solution is rather close in energy a static
stripe phase is quite unlikely to exist in Tl- (Hg-) based cuprates.
It is rather more meaningful to have in mind a picture where
stripes appear as snapshots in the dynamical evolution of the system.
In this context it is interesting to observe that for all
dopings diagonal stripes are higher in energy than the polaron structure.
The low doping phase may resemble that of a 'stripe nematic' 
\cite{kiv98} where orientational symmetry is broken but quantum fluctuations
mainly lead to a restoration of translational symmetry.

\section{Conclusions}
\label{conc}

Within the inhomogeneous GA we have investigated metallic stripe states 
in the one-band extended Hubbard model 
for a fixed value of the on-site repulsion $U/t=10$ as appropriate 
for the cuprates. 
Metallic stripes were shown to be favored by a negative 
$t'/t$ and the physics of this behavior has been clarified in terms of
the competition between the kinetic energy along the stripe and the creation
energy of a stripe. In this regard the HF approximation has been shown to
lead to gross quantitative and often qualitative errors.

Generally the present results are in agreement with our previous 
findings in  the 3-band model,
in particular the behavior of the incommensurability and the
chemical potential, showing that our previous results 
in the 3-band model are quite robust\cite{lor02b}.
On the other hand there are also important differences. 
In the present model SC stripes
are more stable at low doping as compared to BC stripes. 
This is in agreement 
with DMFT computations\cite{fle01} and therefore
seems to be a robust feature of the one-band model and not 
to depend on the approximation. 

One should keep in mind that 
effective models, like the one-band Hubbard model, 
are in principle designed for the description
of electronic excitations in a diluted, uniform phase.
Usually it is not guaranteed that they will give correct results 
when one compares the relative stability of dense non-uniform 
phases.~\cite{bri95} Therefore we consider 
the 3-band result for the stripe state symmetry
as more reliable for the description of real cuprates
 than the one-band calculation.
Further support to this conclusion comes from the 
analysis of transport data.~\cite{nod99,wan01}
Indeed within the  3-band Hamiltonian
where BC stripes constitute the most stable low doping state 
we have explained\cite{lor02b} an anomalous vanishing of the 
transport coefficients\cite{nod99,wan01}
as due to the particular band structure of BC stripes which turns out
to be particle-hole symmetric close to the Fermi level.  
Since the band structure of half-filled SC 
stripes is not particle-hole symmetric around the Fermi level 
(cf. Fig. \ref{bands}) the one-band model cannot account for the
observed vanishing of transport coefficients.~\cite{nod99,wan01}

When the next-nearest neighbor
hopping is fixed to $t'/t=-0.2$ stripes turn out to be half-filled 
at low doping. The behavior of the incommensurability and 
chemical potential is in good agreement with experiment in LSCO. 
Our investigation therefore supports the LDA based one band-mapping
of Ref.~\onlinecite{pav01} where a similar value for the next-nearest
neighbor hopping has been reported for LSCO materials.

As doping increases the energy becomes independent of the 
charge topology.  SC vertical, 
BC vertical, diagonal stripes and a polaron lattice all become degenerate. 
We associate this with a transition from a state with orientational 
order to an isotropic state.~\cite{kiv98} This 
quantum melting of the stripe state  for $x>0.2$
is also reminiscent of the existence of
a quantum critical point in the slightly overdoped regime
as has been proposed by Castellani {\it et al}.~\cite{cas95b}

Since stripes are charged the lateral motion of stripes 
is optically active. Therefore the decrease in the energy gap between
SC and BC stripes should reflect in the   
softening of an optically active electronic mode. 
We have explicitly shown this behavior in a GA+RPA computation 
in the three-band model where the softening of a stripe phason \cite{lor03}
can explain the shift of the mid-infrared
peak in optical conductivity experiments
on LSCO.~\cite{uch91,suz89} We expect a similar behavior in the 
charge response of the one-band model.

Our present results allow for predictions for other cuprate families. 
For larger values of $|t'|$ we obtain lower values of 
$\nu$. Therefore according to Eq.~(\ref{eq:edx}) we predict a steeper curve of 
$\epsilon$ vs. $x$ for small $x$ in Tl and Hg based compounds.
As $|t'|$ increases we also find that the relative stability of stripes 
with respect to isolated polarons decreases. This loss of stability may 
indicate that stripes have a more dynamical character in these materials.

Having in mind the differences and similarities  between the one-band 
and the three-band
description one can still obtain very useful qualitative information
from the one-band description.  
Since within the present work we have set up the relevant parameter
set and stable stripe saddle-points for the LSCO compounds
this represents the first step for an investigation of
the associated charge and spin dynamics.  
Furtheron the one-band model allows for the investigation of
larger clusters and thus the structure of these excitations in q-space
can be investigated in more detail than possible within the
3-band Hamiltonian. Work in this direction is in progress.

\end{document}